\documentclass[pdflatex,sn-nature]{sn-jnl}

\usepackage{graphicx}%
\usepackage{multirow}%
\usepackage{amsmath,amssymb,amsfonts}%
\usepackage{amsthm}%
\usepackage{mathrsfs}%
\usepackage[title]{appendix}%
\usepackage{xcolor}%
\usepackage{textcomp}%
\usepackage{manyfoot}%
\usepackage{booktabs}%
\usepackage[utf8]{inputenc}%
\usepackage{algorithm}%
\usepackage{algorithmicx}%
\usepackage{algpseudocode}%
\usepackage{listings}%
\usepackage{caption}
\usepackage{subcaption}
\usepackage[version=4]{mhchem}

\theoremstyle{thmstyleone}%
\theoremstyle{thmstyletwo}%

\theoremstyle{thmstylethree}%

\providecommand{\nat}{Nature}
\providecommand{\apj}{The Astrophysical Journal}

\providecommand{\mnras}{Monthly Notices of the Royal Astronomical Society}
\providecommand{\pasp}{Publications of the Astronomical Society of the Pacific}

\DeclareUnicodeCharacter{02BB}{\textquoteleft}

\begin{document}

\title[Article Title]{Near-noiseless single-photon detection in the infrared with a megapixel semiconductor array for low-background astronomy}

\author[1,2]{\fnm{Guillaume} \sur{Huber}}\email{ghuber@hawaii.edu}
\author[1]{\fnm{Pavaman} \sur{Bilgi}}\email{pbilgi@hawaii.edu}
\author[2]{\fnm{Charles-Antoine} \sur{Claveau}}\email{caclav@berkeley.edu}
\author[3]{\fnm{Shane} \sur{Jacobson}}\email{shane.jacobson@noirlab.edu}

\author[4]{\fnm{Ian} \sur{Baker}}\email{ian.m.baker@leonardo.com}
\author[4]{\fnm{Daniel} \sur{Owton}}\email{dan.owton@leonardo.com}
\author[4]{\fnm{Vincent} \sur{Isgar}}\email{vincent.isgar@leonardo.com}
\author[4]{\fnm{Chris} \sur{Maxey}}\email{chris.maxey@leonardo.com}
\author[5]{\fnm{Markus} \sur{Loose}}\email{markus.loose@markuryscientific.com}

\author*[1,2]{\fnm{Michael} \sur{Bottom}}\email{mbottom@berkeley.edu}

\affil[1]{\orgdiv{Institute for Astronomy}, \orgname{University of Hawaiʻi at Mānoa}, \orgaddress{\city{Hilo}, \state{HI}, \country{USA}}}

\affil[2]{\orgdiv{Department of Astronomy}, \orgname{University of California, Berkeley}, \orgaddress{\city{Berkeley}, \state{CA}, \country{USA}}}

\affil[3]{\orgdiv{Gemini Observatory}, \orgname{National Optical-Infrared Astronomy Research Laboratory}, \orgaddress{\city{Hilo}, \state{HI}, \country{USA}}}

\affil[4]{\orgdiv{Leonardo UK}, \orgname{Leonardo Corporation}, \orgaddress{\city{Southampton}, \country{UK}}}
\affil[5]{\orgname{Markury Scientific}, \orgaddress{\city{Thousand Oaks}, \orgaddress{\state{CA}}, \country{USA}}}


\abstract{Sensor noise is a fundamental barrier to low-flux astronomical imaging and spectroscopy, where the ideal sensor would unambiguously distinguish individual photon arrivals \cite{rieke2003detection}.  While optical semiconductor arrays can reach single-photon sensitivity, comparable performance in the infrared has largely required superconducting detectors \cite{marsili2013detecting}.  These must be operated at temperatures near absolute zero and require substantially more power, introducing major system-level challenges.  HgCdTe avalanche photodiodes offer a semiconductor alternative by amplifying charge before readout \cite{atkinson2018photon}, but dark current and tunneling effects have limited their use in the faintest regimes \cite{rothman2018physics}.
\\

Here we demonstrate single-photon-resolving operation in a megapixel-format HgCdTe linear-mode avalanche photodiode array operated at conventional cryogenic temperature. Avalanche gain raises photon-induced steps in non-destructive up-the-ramp data above the readout-noise floor while leaving the dark signal largely unamplified, with a false positive rate of approximately 2 false counts per thousand reads per pixel.  These results establish LmAPDs as a promising route toward infrared photon-counting focal planes for future low-background astronomical observatories.
}




\maketitle

\section{Introduction}\label{sec1}


Direct spectroscopy of Earth-like exoplanets is among the most ambitious measurements envisioned for future astronomical observatories. The light reflected by these planets can be ten billion times fainter than that of the stars they orbit, and it must be spectrally dispersed over many pixels to search for molecular absorption features that could reveal the presence of life \cite{2016PASP..128b5003R}. The combination of inherent faintness and dispersion means the photon flux rates are on order of a photon per pixel per hour \cite{2016JATIS...2d1212R}.  Even small amounts of noise can overwhelm such faint signals \cite{borges2024detectability}. The Habitable Worlds Observatory (HWO) will therefore require detectors that approach the ideal limit: detecting individual photons with high efficiency while producing almost no false counts. This requirement is especially challenging in the near-infrared, where detector sensitivity must be reconciled with the practical constraints of a space observatory.

Near-infrared spectroscopy is essential for interpreting potential biosignatures with HWO, providing access to key carbon-bearing molecular lines needed to distinguish biological activity from abiotic alternatives \cite{2025arXiv250714771K}. The scientific reach of sensitive infrared detectors extends far beyond the search for life, from tracing stellar orbits around the supermassive black hole at the Galactic Center to test general relativity \cite{2019Sci...365..664D}, to probing galaxies at cosmic dawn \cite{2024Natur.633..318C}, and uncovering unexpected astrophysical populations such as the compact ``little red dots" \cite{2024ApJ...963..129M}. 

Even from the ground, where larger telescope apertures and fewer constraints on instrument size and power offer greater design flexibility, fine spatial and spectral sampling can leave observations limited by detector noise despite the bright infrared sky. Dark current and read noise limit the sensitivity of iSHELL on NASA’s Infrared Telescope Facility between atmospheric emission lines below 2.5 $\mu$m \cite{2022PASP..134a5002R}, and simulations predict read noise limited performance in several H-band configurations of HARMONI on the forthcoming 39-m Extremely Large Telescope (ELT) \cite{2015MNRAS.453.3754Z}. It has long been recognized that the upcoming class of 30-meter class telescopes, such as the ELT, will not be able to reach their full potential unless infrared sensor read noise is substantially improved \cite{sullivan2012calibrated}.  Reducing detector noise would therefore benefit a broad range of faint-source infrared spectroscopy from both ground and space.

Superconducting detectors are one path to achieving photon-noise-limited imaging. The superconducting energy gap is set by the energy required to break Cooper pairs, orders of magnitude smaller than the $\sim$eV band gaps of semiconductors, making single-photon detection readily achievable. This sensitivity comes with substantial practical constraints. The detectors must be cooled to near absolute zero \cite{you2018superconducting}, and their readout systems can consume orders of magnitude more power than those of conventional semiconductor focal-plane arrays \cite{smith2024mkidgen3, 2026ITAS...36S6750B, 2007SPIE.6690E..0CL}. These cooling and power needs flow down to the instruments, imposing significant system-level challenges, especially for space observatories that have strict mass, vibration, and power constraints. Moreover, while superconducting imagers have reached formats approaching one megapixel \cite{oripov2023superconducting}, the majority have fewer than 1\% as many pixels as conventional semiconductor sensors \cite{lau2023superconducting}.

\begin{figure}[ht]
\begin{center}
\includegraphics[width=0.8\columnwidth]{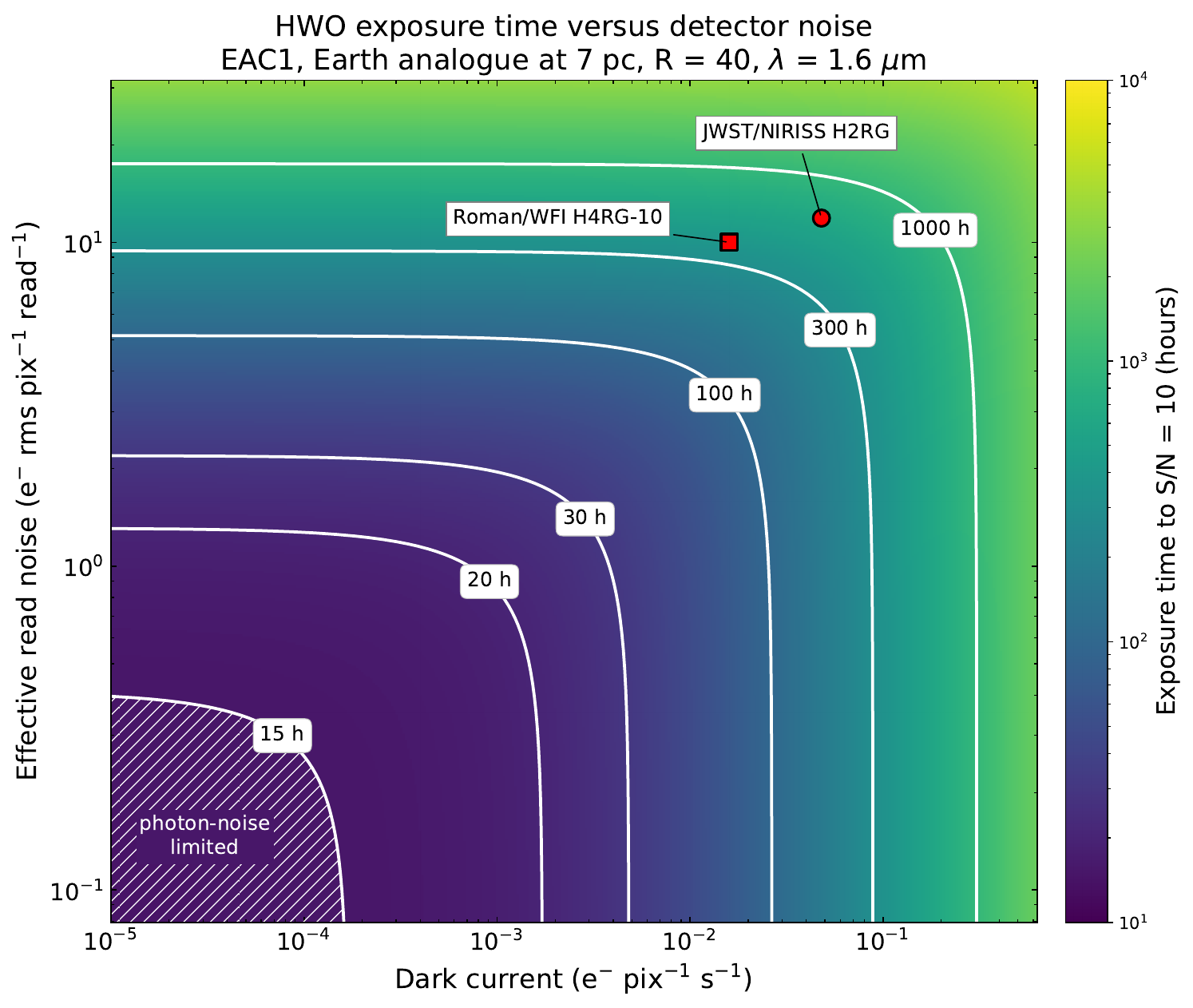}
\caption{\small \textbf{Impact of detector noise on exposure times for HWO exoplanet spectroscopy.} Exposure time required to reach S/N=10 per spectral resolution element near 1.6 $\mu$m with a resolving power R=40 for an Earth-Sun analogue at 7 pc, assuming no residual systematic noise floor. Calculations use pyEDITH \cite{2026arXiv260828477A} with the standard EAC1 observatory configuration, varying only dark current and read noise. Colors and white contours indicate exposure times. The hatched region represents the 15 hours photon-noise limit where detector noise is negligible. Red markers show representative JWST/NIRISS H2RGs (in flight measurements \cite{2023PASP..135i8001D}) and Roman/WFI H4RG-10 (ground-test measurements \cite{2025romanrept..814B}) performance, with read noise expressed on a single-read basis consistent with the pyEDITH noise model; their dark-signal rates include contributions beyond intrinsic dark current.
\label{fig:hwo exp time}
}
\end{center}
\end{figure}

The semiconductor path, by contrast, is well established in infrared astronomy. HgCdTe hybrid arrays, including the state-of-the-art HxRG family of detectors used on the James Webb Space Telescope (JWST) and the Nancy Grace Roman Space Telescope, provide large formats, high quantum efficiency, and excellent stability. However, the combined capacitance of the photodiode, hybrid interconnect and readout input limits charge-to-voltage conversion, increasing the readout noise. The most sensitive infrared arrays exhibit read noise of 10-15 electrons rms per read \cite{2023PASP..135i8001D, 2025romanrept..814B}. Non-destructive readout averaging can reduce this to approximately 5-6 electrons \cite{birkmann2022flight}, still far larger than the $<$0.3 electrons needed to well-resolve individual photon events \cite{fossum2016photon}. Fig.\ref{fig:hwo exp time} illustrates the impact of detector noise on HWO spectroscopy near the 1.6 $\mu$m \ce{CO2} absorption band, which provides important atmospheric context for interpreting potential biosignatures \cite{2026arXiv260426925G}: representative HxRG noise levels imply exposure times of several hundred hours for the example considered here, compared with less than 20 hours in the low-detector-noise regime. Infrared detector noise thus remains a fundamental barrier to low-flux observations, demanding substantial reductions beyond incremental refinement of conventional HxRG arrays, and motivating alternative detector architectures.

Linear-mode avalanche photodiodes (LmAPDs) arrays offer one such route to overcoming this noise floor. In these devices, strong electric fields accelerate photoelectrons sufficiently to produce impact ionization, multiplying the signal before readout and thereby reducing the \textit{effective} readout noise, which can reach subelectron levels \cite{atkinson2018photon}. However, these high fields can also substantially increase dark current through trap-assisted tunneling, in which defect states within the bandgap enable electrons to tunnel into the conduction band without photon absorption. These electrons can undergo avalanche multiplication along with the photoelectrons and produce signals indistinguishable from true photon events, creating false counts. Multiplexer glow, which is infrared photon emission from the readout circuitry \cite{2023AN....34430102P}, is a major contributor to the dark-signal of some HxRG arrays \cite{2020JATIS...6a6001R} and is also observed in LmAPDs \cite{2024SPIE13103E..20H} where glow-induced electrons can undergo avalanche multiplication and contribute false counts. Achieving low-background photon counting therefore requires sufficient avalanche gain to resolve individual photons while keeping detector-induced false events rare.

Here, we present an LmAPD design that achieves a major step towards noiseless infrared photon counting at large format.  The key development is a new graded-bandgap that preferentially amplifies photoelectrons compared to dark signal, while delaying the onset of tunneling current to higher electric fields. Multiplexer glow is also strongly reduced by a metal layer that shields the photosensitive area from photons emitted by the readout electronics. At high avalanche gain, individual photons appear as discrete steps in non-destructive up-the-ramp data, while the effective read noise falls below one electron and the underlying dark signal remains largely unamplified. It is then possible to distinguish photon events from detector dark current and read noise fluctuations with high fidelity, at easily achievable cryogenic operating temperatures of 70 K. These results bring LmAPD arrays closer to a photon-counting regime in which dark current and read noise contribute negligibly to detection error. The false count rate  consists primarily of glow photons that are amplified sufficiently to be miscounted as photoelectrons, we measure this rate as approximately 2 false counts per pixel every 1000 reads. This represents a significant improvement in effective noise compared to current sensors, making it an enabling technology for low-background astronomy.

\section{Overview of LmAPD Physics and Operation}

\subsection{Device Design and Operating Principle}

The LmAPD architecture, developed by Leonardo UK, is based on HgCdTe grown by metal-organic vapor phase epitaxy, enabling the complex heterostructures needed to combine high infrared quantum efficiency, low dark current, and low-noise avalanche multiplication. The detector layer is patterned into mesa-cone diodes and hybridized to the readout integrated circuit (ROIC) by indium bump bonds. The mesa-cone geometry reduces inter-pixel crosstalk and capacitance, and minimizes the volume of material prone to dark current generation \cite{2025JATIS..11a1204B}.  Pixels have a 15 $\mu$m pitch.

\begin{figure}[ht]
\begin{center}
\includegraphics[width=1\columnwidth]{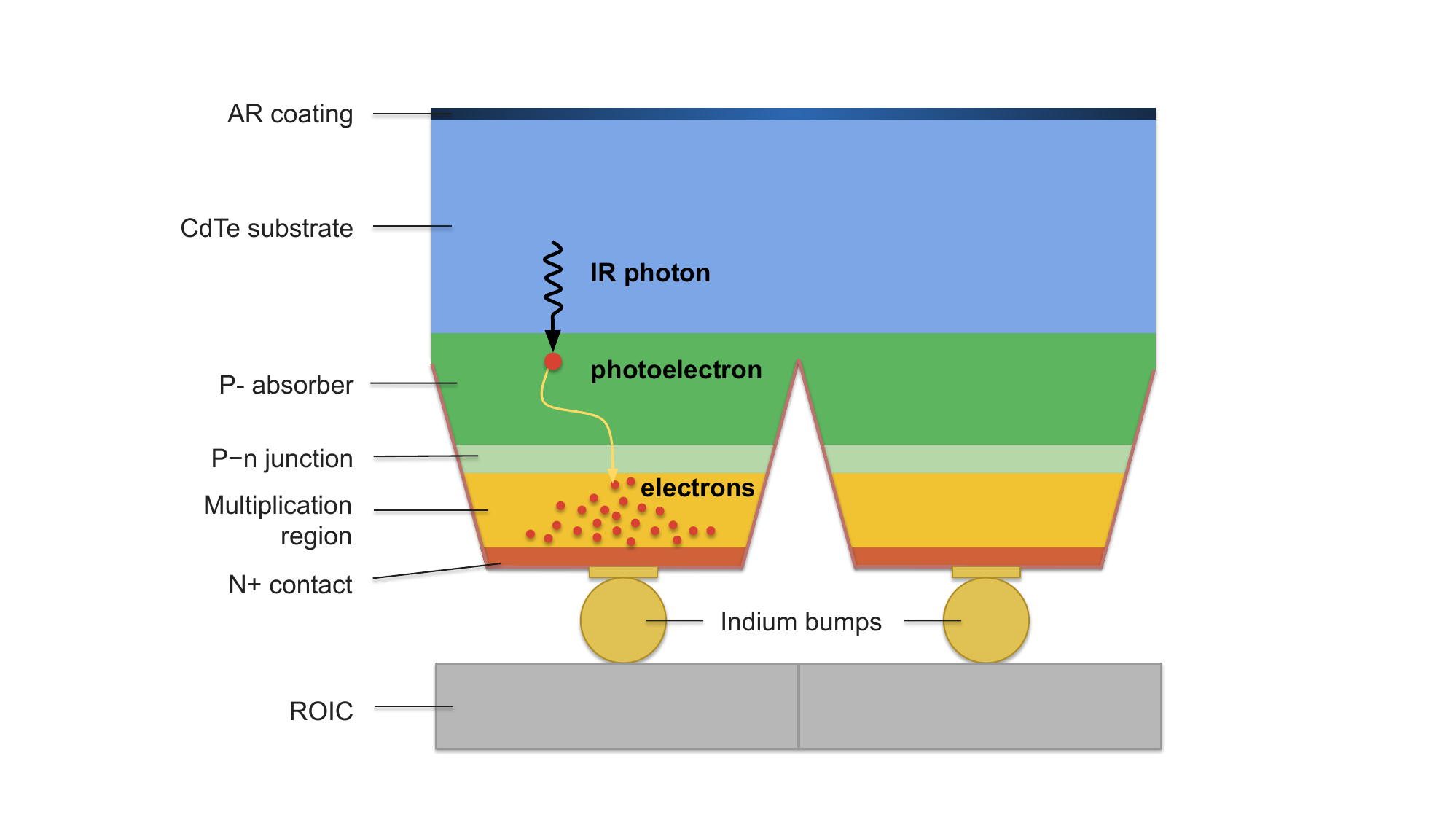}
\caption{\small Schematic of two LmAPD pixels, their architecture and operating principle illustrating the absorption and conversion of a photon into a photoelectron and then its multiplication in the avalanche region.
\label{fig:lmapd}
}
\end{center}
\end{figure}

The physical operating principle by which a photon is absorbed and multiplied by a pixel, resulting in a measurable output, is illustrated in Fig.\ref{fig:lmapd}. The near-infrared photon converts to an electron via the photoelectric effect in the HgCdTe absorber, which drifts into the high-field multiplication region. There, the electron initiates avalanche cascading, generating secondary electrons through impact ionization. The resulting charge packet is then collected and read out by the ROIC. The design presented here has a spectral response bounded at short wavelengths by absorption in the CdTe layer ($<$0.8 $\mu$m) and extends to 2.5 $\mu$m, although that cutoff may be tuned though cadmium fraction as in conventional HgCdTe engineering.

\subsection{APD Gain and Readout Noise}

HgCdTe is ideal for avalanche multiplication as it provides a strong asymmetry between electron and hole transport \cite{2011OptLT..43.1358S}: heavy holes have low mobility and efficiently lose energy to optical phonon scattering, while electrons have a small effective mass, high mobility at cryogenic temperatures, and weak optical phonon scattering. Electrons can therefore gain sufficient energy from modest applied fields to initiate impact ionization and thus avalanche cascading.

Because the avalanche process occurs in the diode before the multiplexer readout, the photo-electrons are amplified while the downstream readout noise remains the same. Referred to the photon input, the read noise therefore decreases in inverse proportion to the avalanche gain $M$:

\begin{equation}
    \sigma_{read, \, input} = \sigma_{read, \,output}/M \, .
    \label{DC eq}
\end{equation}

At sufficiently high gain, the photon signal can be amplified above the read noise floor, thus opening the path for single photon detection.

\subsection{Dark Current and Tunneling Limits}

Dark current in LmAPDs occurs through two main mechanisms: thermal generation across the bandgap, and trap-assisted tunneling in high-field regions. Those effects can be reduced by varying the cadmium fraction during growth, which allows bandgap widening and grading across the multiplication region. In particular, this confines the dark current region away from the absorber so that dark current electrons experience minimal avalanche gain.

\begin{figure}[ht]
\begin{center}
\includegraphics[width=1\columnwidth]{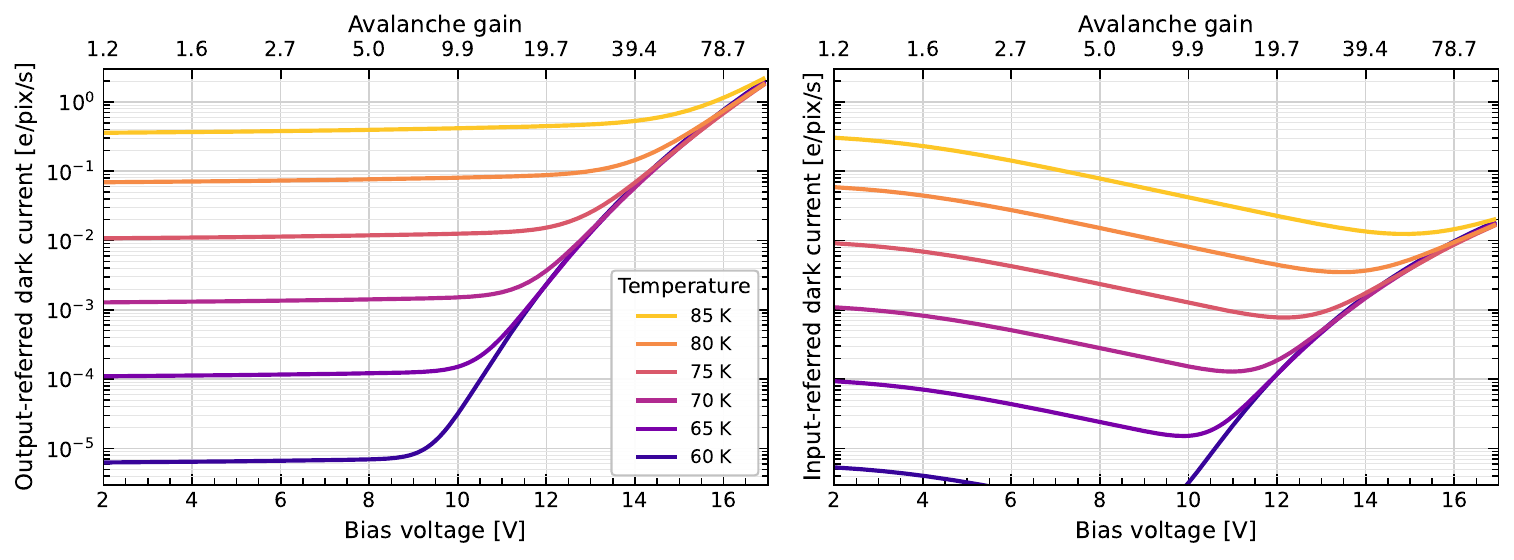}
\caption{\small Model of the output and input referred dark current as a function of bias voltage and avalanche gain for different detector temperatures.
\label{fig:DC}
}
\end{center}
\end{figure}

Fig.\ref{fig:DC} shows the dependence of dark current on temperature, bias voltage, and avalanche gain. At a fixed temperature, the output dark signal remains nearly constant over most of the bias range, increasing only at higher bias where tunneling becomes the main contributor. For an exposure generating $N_{DC}$ dark current electrons, the input-referred dark current noise contribution is

\begin{equation}
    \sigma_{DC, \, input}^2 = N_{DC}(M)/M^2\, .
    \label{DC eq}
\end{equation}

This expression initially decreases with avalanche gain, before rising again once tunneling dominates.

\subsection{Multiplexer Glow}

Multiplexer glow is a general term for electroluminescence from the silicon circuitry. It is associated with MOSFETs in the saturated regime, where high fields can be generated at the drain end of the channel. Electroluminescence from the pixel circuitry is particularly important as it can interfere with non-destructive readout strategies. Mitigation in recent devices has involved MOSFET geometry, bias voltage, operating currents, and dense metal shielding. These changes have resulted in a significant reduction in the glow level to approximately 0.01 e-/pixel/frame \cite{2024SPIE13103E..20H} in Ike Pono, the latest megapixel iteration and the array studied here. Preliminary measurements indicate that glow photons are mainly absorbed within the multiplication region and that the resulting photo-electrons, like dark current electrons, do not experience the full avalanche gain. In future design iterations, it is anticipated that electroluminescence can be fully eliminated by avoiding the deep saturation regime of critical MOSFETs.

\subsection{Readout strategies for LmAPDs}

LmAPDs support non-destructive readout, allowing the same sampling schemes commonly used with conventional infrared detector arrays. A simple and common method for measuring the accumulated signal is correlated double sampling (CDS), which takes the difference between non-destructive reads at the beginning and end of an exposure. Fowler sampling extends this approach by averaging groups of reads at the beginning and end, reducing the read noise contribution. However, glow generated by the additional reads introduces accumulating shot noise, setting a noise floor \cite{2026arXiv260800354B}. Another widely used readout scheme is up-the-ramp sampling, in which non-destructive reads are acquired throughout the exposure, the signal is then estimated by fitting a slope to the accumulating charge \cite{2024PASP..136d5004B}.

In this paper, we use rapid up-the-ramp sampling to resolve individual photon events directly. This method preserves the temporal evolution of each pixel, providing access not only to the number of detected photons but also to their arrival time, while revealing time-dependent effects such as dark current drifts, persistence, or cosmic-ray events. Beyond measuring an integrated signal, this approach demonstrates the stronger and more physically relevant capability required for low signal imaging: resolving individual photon events directly in the time domain.

\section{Experimental Setup and Results}

\subsection{Laboratory Setup}

\begin{figure}[ht]
\begin{center}
\includegraphics[width=1.0\columnwidth]{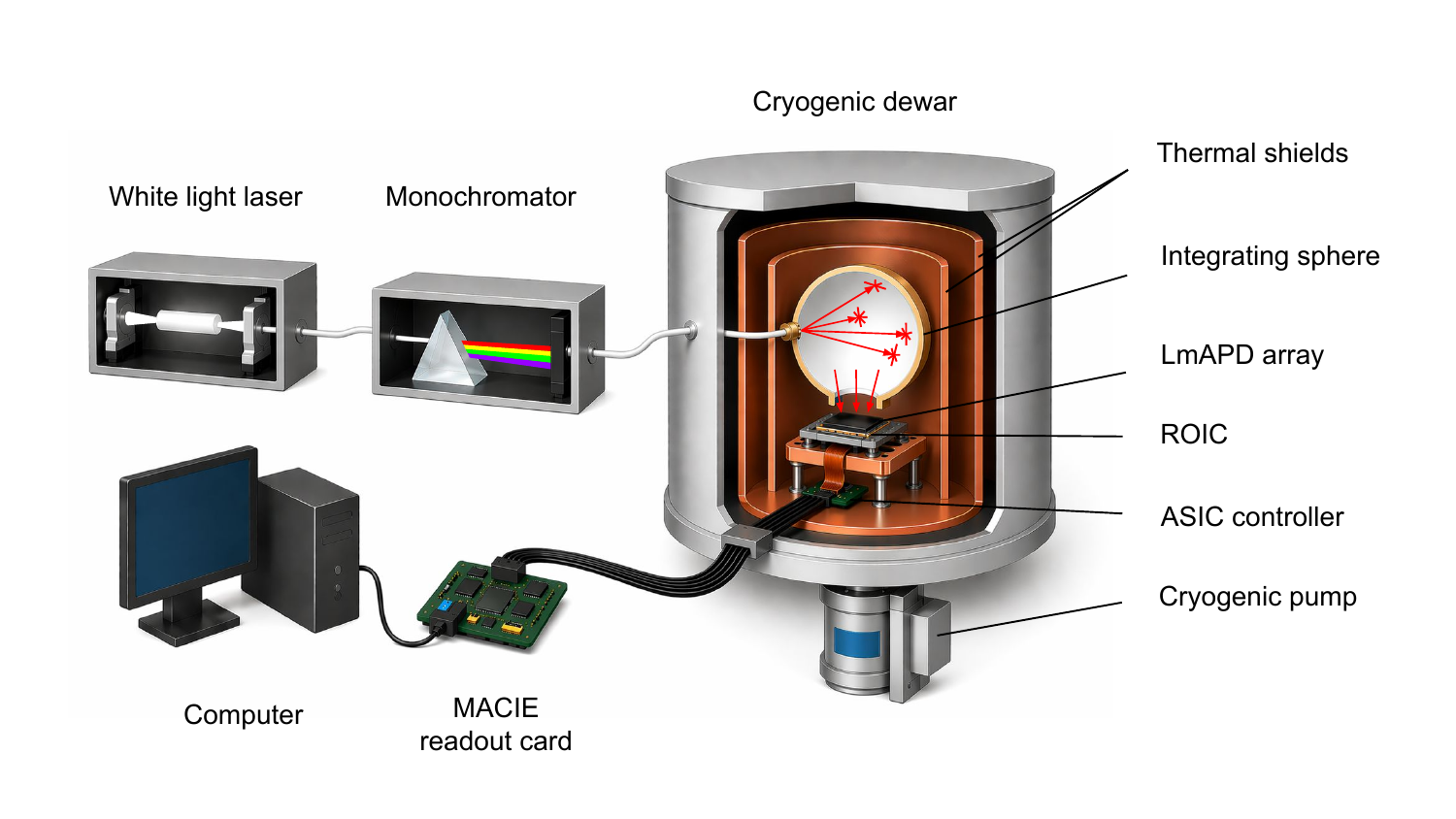}
\caption{\small Illustration of the LmAPD laboratory setup.
\label{fig:lab setup}
}
\end{center}
\end{figure}

Our experimental setup is illustrated in Figure \ref{fig:lab setup}. The test chamber consists of a cryo-cooled vacuum dewar with milliKelvin stability and a pressure below $10^{-6}$ Torr. The illumination source is a super-continuum  white light laser with a variable repetition rate, providing a tunable photon flux with sub-percent stability over a dynamic range of 200. Before the cryostat, a monochromator with a 5 nm bandwidth allows for wavelength selection. The beam then enters the cryo-chamber via optical fiber and passes through an integrating sphere, producing a flat field on the detector with a thermal background below 4 photon/pixel/day. In the present measurements, the detector was cooled to 70 K and illuminated with 1500 nm light.

The detector is a $1024 \times 1024$ pixel LmAPD array controlled by a SIDECAR ASIC \citep{2003SPIE.4841..782L}, a compact micro-controller that both operates the detector and digitizes its output. It connects to the control computer through a MACIE (Multi-purpose ASIC Control and Interface Electronics) board. Two external power supplies provide the voltages for the detector and its electronics. Further details on the readout chain and data acquisition can be found in \citep{2024arXiv241109185C}.

\subsection{Single Photon Detection}

\begin{figure}[ht]
\begin{center}
\includegraphics[width=0.8\columnwidth]{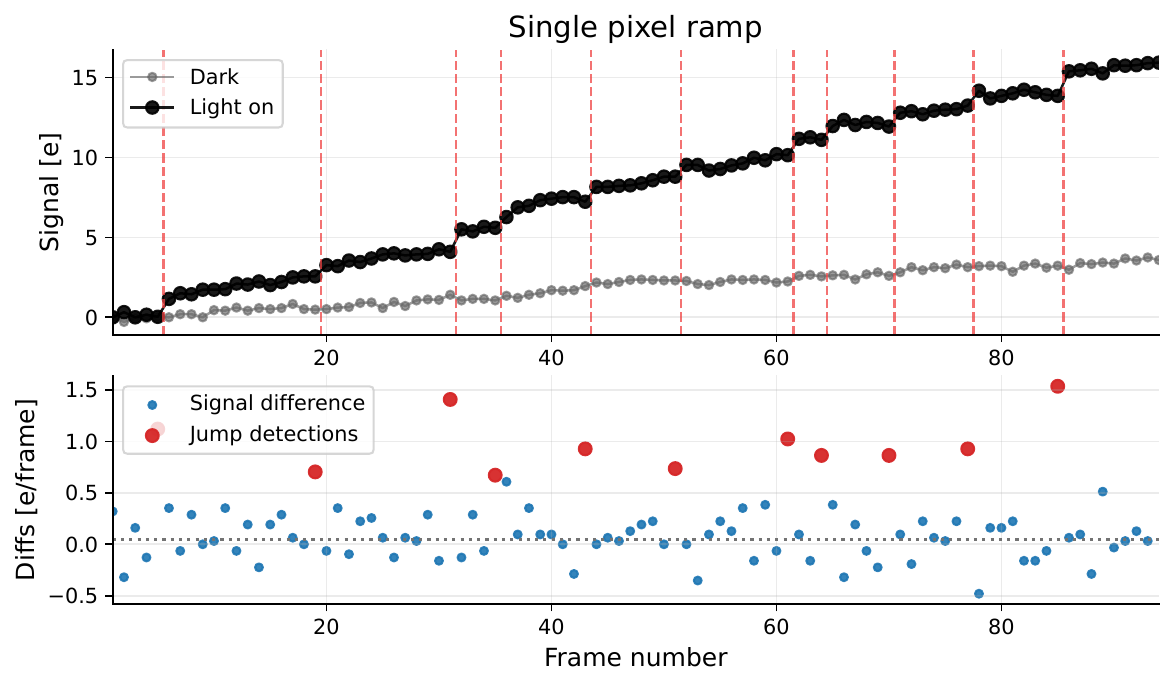}
\caption{\small \textit{Top:} at high gain, photon events manifest as steps in the data, highlighted in red. Displayed is also dark data for the same pixel where no jumps appear. \textit{Bottom:} frame-to-frame signal difference, showing how photon-induced jumps exceed the readout noise.
\label{fig:photon jumps}
}
\end{center}
\end{figure}

Under low-illumination conditions, LmAPDs operating at high avalanche gain can resolve individual photon events. Fig.~\ref{fig:photon jumps} shows a 66 second up-the-ramp sequence from a single pixel, acquired at 1.45 frames/second, in which photon detection events appear as discrete steps. For this measurement, the detector was operated at an avalanche gain of 78.7, corresponding to an effective readout noise of 0.45 e- rms. While consecutive non-destructive reads still exhibit scatter from readout noise, the high avalanche gain amplifies the photonic signal sufficiently for it to stand above this noise floor. In contrast, an equivalent measurement with the light source off shows only the smooth increase from largely unamplified dark current and glow, with no discontinuities from photons. As such, dark current and glow do not significantly affect single photon detection in those operating conditions.

Rather than using the up-the-ramp sequence conventionally to estimate the incident flux from a fitted slope, we use the individual reads to identify photon events directly in the time domain. This non-destructive readout method preserves the temporal evolution of each pixel, providing access not only to the number of detected photons but also their arrival times.

While a thresholding algorithm can detect most jumps, we adopt a likelihood-based framework originally developed for cosmic-ray rejection in up-the-ramp detector data \cite{2024PASP..136d5005B}. The method evaluates whether omitting an individual read difference produces a significant improvement in the likelihood of a covariance-weighted ramp fit, thereby identifying discrete jump events directly in the time domain. In contrast to fixed amplitude thresholding, this approach accounts for the expected noise model from the up-the-ramp readout.

\subsection{Flux-Dependent Jump Count at the Array Level}

\begin{figure}[ht]
\begin{center}
\includegraphics[width=0.8\columnwidth]{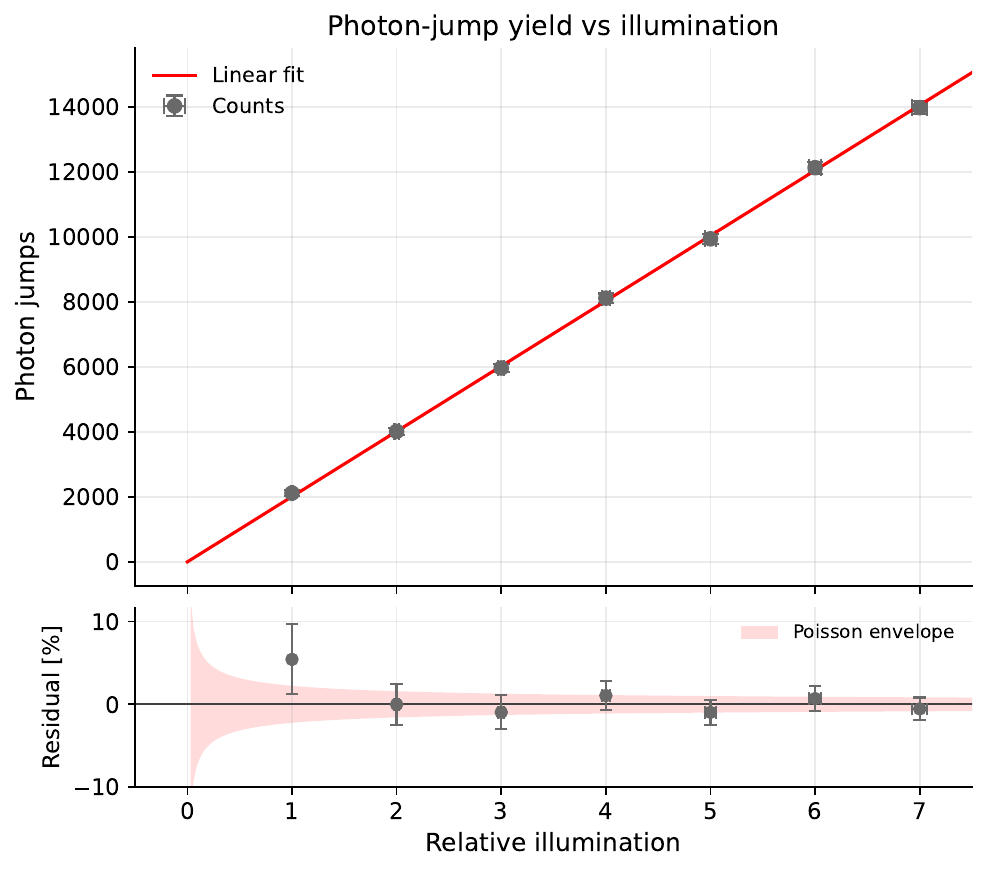}
\caption{\small \textbf{Flux dependence of detected photon jumps.} Number of detected jumps in 1000 selected pixels as a function of illumination level. The red line shows a linear fit, and the lower panel shows fractional residuals relative to the fit. Vertical error bars show bootstrap uncertainties from resampling the pixel population, horizontal uncertainties from the illumination calibration are approximately 1\%. The shaded region indicates the expected Poisson counting envelope, $\pm \sqrt{N_{fit}}/N_{fit}$, for an ideal photon-limited jump-counting measurement.
\label{fig: linear fit}
}
\end{center}
\end{figure}

By analyzing a large number of pixels, we can investigate the general ability of the array to perform single photon detection. In Fig.\ref{fig: linear fit}, we randomly select 1000 pixels among the 50\% of the array with the lowest dark levels and run our jump-detection algorithm on 300-frame ramps at 1.45 frames/s. We repeat the experiment spanning seven illumination levels using our tunable light source and find that the number of jumps detected increases proportionally with incident flux. This confirms that the observed events are associated with the absorption of signal photons, rather than intrinsic detector artifacts or external effects.

To estimate the false-positive rate of our approach, we also acquire dark ramps with the illumination source blocked (\textit{e.g.} gray data in Fig.\ref{fig:photon jumps}) and process them with the same jump-detection algorithm. For the 300-frame ramps used here, we detect 0.65 jumps per pixel per ramp on average, corresponding to a detected dark-event rate of $2.2\times10^{-3}$ events/pixel/frame. We attribute those rare threshold-crossing events primarily to glow-induced jumps and subtract this small offset from the data in Fig.\ref{fig: linear fit}. After correction, the remaining jumps scale linearly with illumination level, confirming that the flux-dependent signal is photon-induced and that the detector and the measurement method remain linear over the tested range.

\section{Conclusion and Future Efforts}

The megapixel-scale infrared devices presented in this paper have demonstrated the capability to detect single-photon events at conventional cryogenic temperature, a realm formerly limited to superconducting sensors.  Together with the good material properties of HgCdTe such as high infrared quantum efficiency and manufacturability, the remaining challenges lie mainly in process control for defects, uniformity, and persistence.  The operating mode presented in this paper, where the sensors are read out up-the-ramp with jumps identified as photon events, offers a path towards noiseless infrared photon-number resolution without the operability challenges of superconductors.  Maturing such detectors could extend the sensitivity of low-signal astronomical imaging and spectroscopy, particularly for observations of faint sources such as Earth-like exoplanets, high-redshift galaxies, and supernovae at cosmological distances.

Future efforts are focused on scaling devices to larger formats while preserving high uniformity and low defect levels, with particular emphasis on bandgap engineering to further reduce tunnel currents.  In particular, we are currently designing 3-side buttable $2k\times 2k$ arrays that will feature faster readout (up to 60 frames/second) and more efficient differential readout functionality, which will in effect reduce the integrated dark current per frame, with a goal of photon counting capability in $\sim$100\% of pixels.  A sensor of this format would be well matched to modern astronomical instruments, which mainly use $2k\times 2k$ or larger formats, and a transformative technology for the Habitable Worlds Observatory and future 30-meter class telescopes.

\clearpage

\section{Methods}

\subsection{Fabrication}

The LmAPD diodes are grown by metal-organic vapor-phase epitaxy (MOVPE), a technology that enables the growth of complex heterostructures with the bandgap and doping independently controlled on a 0.1 $\mu $m scale. This allows the absorbers, junctions, and the avalanche region to be optimized separately.

These layers are grown on GaAs substrates, which are available as high-quality, epi-ready single-crystal wafers in large formats. Current fabrication methods can process 100-mm-diameter wafers, each accommodating multiple large-format detector arrays. GaAs is removed by a selective etch after hybridization to remove thermal stresses and internal reflections. To accommodate the 14\% lattice mismatch of GaAs with HgCdTe,  buffer layers are used to mitigate misfit dislocations. Then, in the remaining material, defect-revealing etches show only a faint pattern of etch pits from weak mosaic crystal boundaries. Threading dislocations in MOVPE-grown arrays are strongly reduced compared with LPE grown on $<$111$>$B-oriented CdZnTe substrates, so MOVPE arrays tend to exhibit low levels of defective pixels.

\begin{figure}[ht]
\begin{center}
\includegraphics[width=0.5\columnwidth]{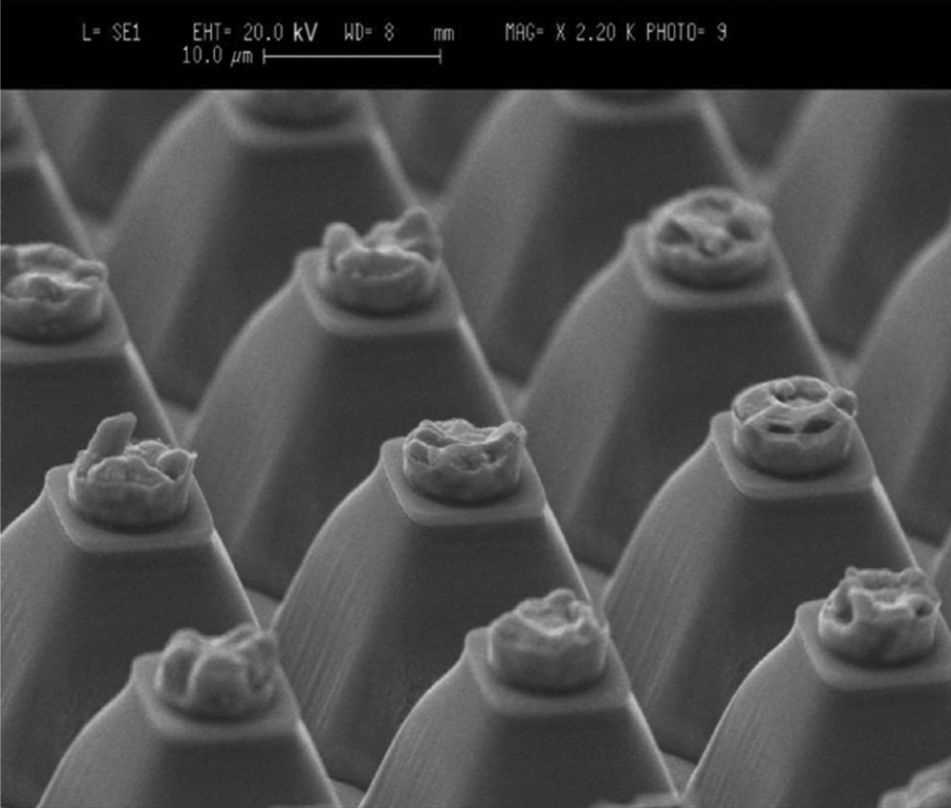}
\caption{\small Deep slot mesa structure with the indium bump interconnect
\label{fig: mesa indium}
}
\end{center}
\end{figure}

The detector arrays are fabricated using wafer-scale processing. Each pixel is electrically isolated by a mesa slot, shown in Fig.\ref{fig: mesa indium}, that extends through the absorber and suppresses lateral carrier collection and blooming. It also reduces optical crosstalk and inter-pixel capacitance while preserving the modulation transfer function. One advantage of HgCdTe is that it permits hetero-passivation: the mesa sidewalls are coated with a CdTe layer that is inter-diffused at high temperature. The widening of the bandgap around the edges of the absorber effectively separates carriers from surface states and minimizes junction currents where the junction intercepts the sidewall. Hetero-passivation is important for achieving uniform detector performance, reproducible production yields, and long-term device stability.

The photodiode array is subsequently hybridized to the ROIC via indium bump bonding to form the complete focal-plane assembly. An indium bump located at the tip of each pixel mesa connects the photodiode N+ contact pad to the corresponding ROIC input, providing an independent electrical interconnect for every pixel across the $1k\times 1k$ array.

\subsection{Likelihood-based jump detection}\label{secA1}

The photon jump detection algorithm is adapted from a likelihood-based approach developed for non-destructively read detectors \cite{2024PASP..136d5005B}. The original framework was designed to identify cosmic ray jumps in up-the-ramp data containing a steady count-rate background with Poisson and readout noise. This is directly analogous to the objective of detecting discrete single-photon jumps superimposed on a steady dark background. The advantages of this method are that it provides the optimal linear fit under the assumed Poisson and readout noise covariance model, and that its computational cost scales linearly with the number of reads, which is important when processing large datasets.

The cosmic ray algorithm can be divided into two components:
\begin{enumerate}
    \item deriving the covariance model of the data;
    \item likelihood-based fitting and jump identification.
\end{enumerate}
For our purposes, the second part is essentially unchanged. We therefore describe only the adaptation of the covariance model for the single-photon detection experiment. For consistency, we keep the notation as similar as possible to the aforementioned work.

The general goal of the fit is to minimize the chi-square statistic

\begin{equation}
    \chi^2 = \left(\mathbf{N}_{\mathrm{meas}}-\mathbf{N}_{\mathrm{model}}\right)^T
\mathbf{C}^{-1}
\left(\mathbf{N}_{\mathrm{meas}}-\mathbf{N}_{\mathrm{model}}\right)
\end{equation}

where $\mathbf{N}_{\mathrm{meas}}$ is the vector of scaled differences between consecutive resultants, or frame groups, $\mathbf{N}_{\mathrm{model}}$ is the corresponding constant-count-rate model estimate, and $\mathbf{C}$ the covariance matrix of the dataset which encapsulates the shot and readout noise:

\begin{equation}
    \mathbf{C} = a\mathbf{C}_{\gamma} + \sigma^2\mathbf{C}_r
\end{equation}

where $a$ is the background count rate, $\sigma^2$ is the readout noise variance, and $\mathbf{C}_{\gamma}$ and $\mathbf{C}_{r}$ are the shot and readout noise covariance matrices, respectively.

The covariance matrix of the resultant differences is tridiagonal

\begin{equation}
    \mathbf{C} =
    \begin{bmatrix}
    \alpha_1 & \beta_1 & 0 & 0 & \cdots \\
    \beta_1 & \alpha_2 & \beta_2 & 0 & \cdots \\
    0 & \ddots & \ddots & \ddots & \\
    0 & \cdots & \beta_{n-2} & \alpha_{n-1} & \beta_{n-1} \\
    0 & \cdots & 0 & \beta_{n-1} & \alpha_n
    \end{bmatrix}
\end{equation}

where

\begin{equation}
    \begin{aligned}
\alpha_i &=
\sigma^2
\left(\frac{1}{(\Delta_i t)^2}\right)
\left(\frac{1}{n_i}+\frac{1}{n_{i+1}}\right)
+a
\left(\frac{1}{(\Delta_i t)^2}\right)
\left(\tau_i+\tau_{i+1}-2\langle t_i\rangle\right),
\\[1em]
\beta_i &=
\sigma^2
\left(\frac{1}{(\Delta_i t)(\Delta_{i+1} t)}\right)
\left(\frac{-1}{n_{i+1}}\right)
+a
\left(\frac{1}{(\Delta_i t)(\Delta_{i+1} t)}\right)
\left(\langle t_{i+1}\rangle-\tau_{i+1}\right).
\end{aligned}
\end{equation}

Here, $n_i$ is the number of nondestructive reads averaged into the resultant $i$, $\langle t_i \rangle$ is the mean acquisition time of that resultant, and $\tau_{i}$ is the variance-weighted effective time defined by Brandt as

\begin{equation}
    \begin{aligned}
    \tau_i = \frac{1}{n_i^2}\sum_k \left(2n_i + 1 - 2k\right)t_{ik}
\end{aligned}
\end{equation}

where $t_{ik}$ is the acquisition time of read $k$ within resultant $i$.

In our data, each resultant consists of a single non-destructive read acquired at constant cadence. Thus, $n_i=1$, $\tau_i = \langle t_{i}\rangle = t_i$, and $\Delta_i t = \Delta t$. The adapted covariance coefficients reduce to

\begin{equation}
    \begin{aligned}
\alpha_i &=
\frac{a}{\Delta t}
+
\frac{2\sigma^2}{\Delta t^2},
\\[1em]
\beta_i &=
-\frac{\sigma^2}{\Delta t^2}.
\end{aligned}
\end{equation}

The Poisson contribution is therefore confined to the diagonal because charge accumulated during distinct frame intervals is independent, whereas adjacent differences remain negatively correlated through the readout noise of the frame shared between them.

Unlike the original implementation, which estimates the baseline count rate $a$ from the ramp being analyzed, we use a reference dark ramp to define the baseline slope model so that the statistic is optimized for departures from the dark current and glow background rather than from the illuminated signal. We also measure the readout noise $\sigma$ directly from the dark ramp after subtracting its linear fit. This makes the covariance matrix more appropriate because it is based on the physically relevant null model: uncontaminated detector data acquired with the illumination source blocked.

The remainder of the cosmic ray detection algorithm is unchanged. For each candidate event location, it compares the covariance-weighted likelihood of the ramp using all valid differences with that obtained after omitting one difference, or two adjacent differences where appropriate. The candidate producing the largest statistically significant reduction in $\chi^2$ is masked, the constant baseline rate is refitted using the remaining differences, and the procedure is repeated iteratively until no additional candidate exceeds the adopted likelihood threshold.

\subsection{HWO exposure-time simulations}

We used pyEDITH \cite{2026arXiv260828477A} to calculate the exposure time required to reach a signal-to-noise ratio of 10 per spectral resolution element near 1.6 $\mu$m for an Earth–Sun analogue at 7 pc, representative of an illustrative observing scenario in the search for biosignatures on nearby Earth-like exoplanets with HWO. The stellar spectrum was represented by a 5800 K blackbody with a solar radius, and the planet-to-star flux ratio was taken from the Earth reflected-light template distributed with pyEDITH. We adopted a projected separation of 1 au (0.143 arcsec), an exozodiacal dust level of one zodi, and an absolute ecliptic latitude of 30° for the local zodiacal background. Calculations used the EAC1 observatory configuration in integral-field spectroscopy mode, with a resolving power of R = 40, the minimum resolution to avoid a degeneracy between \ce{CO2} and \ce{CO} in atmospheric retrievals \cite{2026arXiv260426925G}. Dark current and single-read noise were varied independently, while all other detector parameters, including the read cadence, were held fixed at their EAC1 values. Residual systematic noise floors from stellar leakage and exozodiacal-light subtraction were set to zero while retaining their photon-noise contributions. We therefore evaluate detector requirements assuming that the coronagraphic residuals and exozodiacal background can be calibrated to the photon-noise limit.

Representative HxRG measurements were overlaid for comparison. For JWST/NIRISS \cite{2023PASP..135i8001D}, the adopted correlated-double-sampling noise of 16.8 e- was divided by $\sqrt{2}$ to obtain a single-read equivalent of 11.9 e- rms, assuming independent read errors. The measured dark signal of 0.515 e-/pixel/frame was divided by the 10.737 s/frame time, giving an equivalent rate of 0.048 e-/pixel/s. For Roman/WFI \cite{2025romanrept..814B}, we adopted a dark-signal rate of 0.016 e-/pixel/s and a single-read noise of 10 e- rms. These dark-signal measurements include instrument-dependent contributions beyond intrinsic detector dark current such as glow and thermal background. The comparison therefore illustrates the exposure-time consequences of the adopted noise values within a common HWO configuration.

\clearpage

\backmatter

\bmhead{Supplementary information}

If your article has accompanying supplementary file/s please state so here. 

\bmhead{Acknowledgements}
The authors acknowledge Donald N. B. Hall (1944-2020), who first understood the potential of LmAPDs for low-flux imaging, and initiated this line of research.  This work was supported by the National Aeronautics and Space Administration (NASA) Strategic Astrophysics Technology (SAT) program under grants 80NSSC20K0330 and 80NSSC25K7007.

\section*{Declarations}
The authors declare the following competing interests: IB, DO, VI, and CM are employees of Leonardo UK, the manufacturer of the array technology that is the subject of this study. Contractual non-disclosure agreements guarantee that all data publication remains solely at the discretion of the academic partners; as such, the company had no role in the underlying study design, data analysis, or oversight regarding the decision to publish.  The remaining authors declare no competing interests.

\end{document}